\documentclass[11pt,twoside]{article}
\pdfoutput=1
\usepackage{verbatim}
\usepackage{amssymb}
\usepackage{amsfonts}
\usepackage{amsmath}
\usepackage{pifont}
\usepackage[spanish,english]{babel}
\usepackage{subfig}
\usepackage{placeins}
\usepackage[pdftex]{graphicx}
\usepackage{verbatim}
\usepackage{bm}
\usepackage{longtable}
\usepackage{fancyhdr}
\usepackage{placeins}
\usepackage[colorlinks=true,linkcolor=blue,urlcolor=blue,citecolor=blue,pdftex]{hyperref}

\begin{document}

\fancypagestyle{plain}{%
\fancyhf{}%
\fancyhead[LO, RE]{XXXVIII International Symposium on Physics in Collision, \\ Bogot\'a, Colombia, 11-15 september 2018}}

\fancyhead{}%
\fancyhead[LO, RE]{XXXVIII International Symposium on Physics in Collision, \\ Bogot\'a, Colombia, 11-15 september 2018}

\title{Higgs boson measurements at the LHC}
\author{Guillaume Unal $\thanks{e-mail: Guillaume.Unal@cern.ch}$,
on behalf of the ATLAS and CMS Collaborations $\thanks{Copyright 2018 CERN for the benefit of the ATLAS and CMS Collaborations. CC-BY-4.0 license.}$ \\
\\ CERN }
\date{}
\maketitle

\begin{abstract}
Measurements of the Higgs boson production and decay performed at the Large Hadron Collider by the
ATLAS and CMS experiments are reviewed. These measurements are based on proton-proton collision
data at $\sqrt{s}$=~13~TeV, corresponding to integrated luminosities ranging from 35 to 80~fb$^{-1}$.
With these datasets, the associated production of the Higgs boson with a $t\bar{t}$ pair is observed and the
decay of the Higgs boson to $b\bar{b}$ pairs is established. Measurements involving leptonic and bosonic final
states are described. The combined constraints on the Higgs boson coupling properties are summarized.
\end{abstract}

\section{Introduction}

After the discovery~\cite{atlas-discovery,cms-discovery}
of a particle consistent with the Standard Model (SM) Higgs boson ($H$) at the Large
Hadron Collider (LHC) in 2012 by the ATLAS~\cite{ATLAS-exp} and CMS~\cite{CMS-exp} experiments, the properties of this particle
were studied using the full run 1 (2011-2012) dataset and found to be consistent with the
Standard Model expectations: The mass was measured with a 0.2\% accuracy~\cite{mass-run1}, 
all tested alternatives
to the $0^+$ spin-parity assignment were rejected~\cite{atlas-cp-run1,cms-cp-run1}
and the couplings were found to be consistent
with the SM predictions with accuracies reaching around 10\% in the most favorable cases~\cite{couplings-run1}.
However, only the gluon fusion and VBF production processes and 
only the decays to bosons were clearly observed, with the decays to $\tau\tau$ pairs observed
at the five standard deviations level only with the combination of ATLAS and CMS results. Direct observations
of the coupling to $b$ and $t$ quarks were lacking, the later being only inferred from the gluon fusion loop induced
Higgs production.

The run 2 data taking period started in 2015 at an increased centre of mass energy of 13~TeV and increased
instantaneous luminosity, reaching $2 \cdot 10^{34}$cm$^{-2}$s$^{-1}$, twice the LHC design luminosity.
Among the main goals of this data taking period were the discovery of the $t\bar{t}H$ production process,
the observation of the decay $H \rightarrow b\bar{b}$ and precise measurements of the $H$ boson properties
in the decay channels involving bosons ($WW^*$,$ZZ^*$ and $\gamma\gamma$).
The results presented in this paper are based on either data collected in 2015-2016, corresponding to an integrated
luminosity around 35~fb$^{-1}$ or on data collected between 2015 and 2017, corresponding to an integrated luminosity
around 80~fb$^{-1}$.

At the LHC, the main production mode for the Higgs boson is the gluon fusion process, mostly mediated
by a top quark loop. The predicted cross-section for a Higgs boson mass of 125~GeV is about 50~pb.
Thanks to recent N3LO QCD computations~\cite{n3lo}, the accuracy on the predicted cross-section is about 5\%.
The second most important production mode, with a cross-section of 3.7~pb, is the vector boson
fusion process (VBF) which has a distinct signature of two scattered quarks in addition to the produced
Higgs boson. The associated production of a Higgs boson with a vector boson $W$ or $Z$ ($VH$ production)
has a predicted cross-section of about 2~pb. Finally, the $t\bar{t}H$ production, which is sensitive at tree
level to the coupling between top quarks and $H$, has a predicted cross-section of 0.5~pb. This production
mode benefits most from the centre of mass energy increase from 8~TeV to 13~TeV with a factor four increase
in the predicted cross-section.
The dominant
decay mode of the Higgs boson, with a predicted branching ratio of 58\%, is to $b\bar{b}$ pairs. The other decay modes are
$\tau\tau$ (6.2\%), $c\bar{c}$ (2.9\%), $\mu\mu$ (0.02\%), $WW^*$ (21\%), $ZZ^*$ (2.6\%), $gg$ (8.2\%), 
$\gamma\gamma$ (0.2\%) and $Z\gamma$ (0.15\%).
Details about the predicted cross-section and branching ratio values can be found in Ref.~\cite{cross-sections} and references
therein.

\section{Observation of the $t\bar{t}H$ production}

\begin{figure}[htb]
\centering
\begin{minipage}{0.45\textwidth}
  \subfloat[]{\includegraphics[scale=0.32]{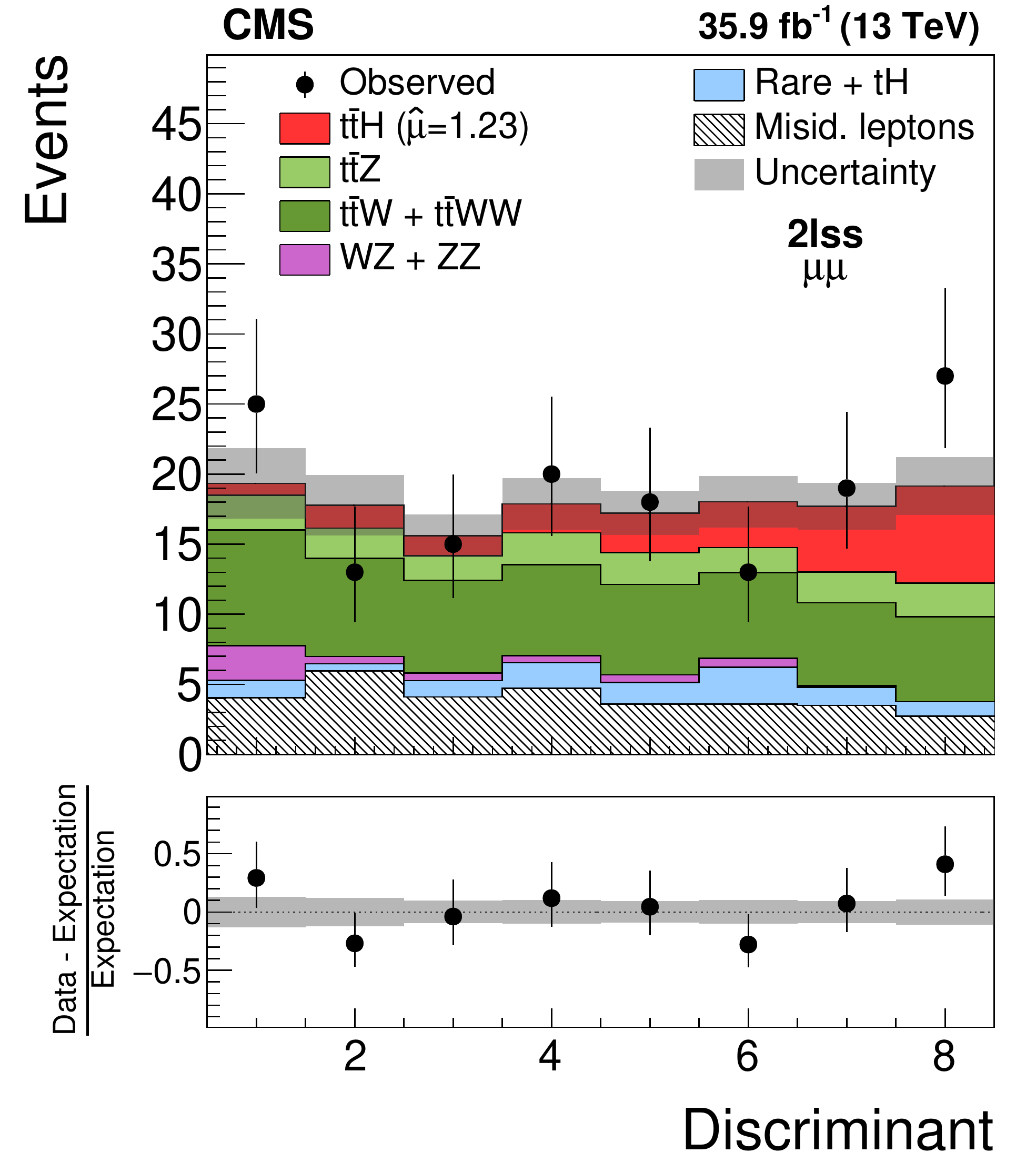}\label{fig:ttH_a}}
\end{minipage}
\begin{minipage}{0.45\textwidth}
  \subfloat[]{\includegraphics[scale=0.43]{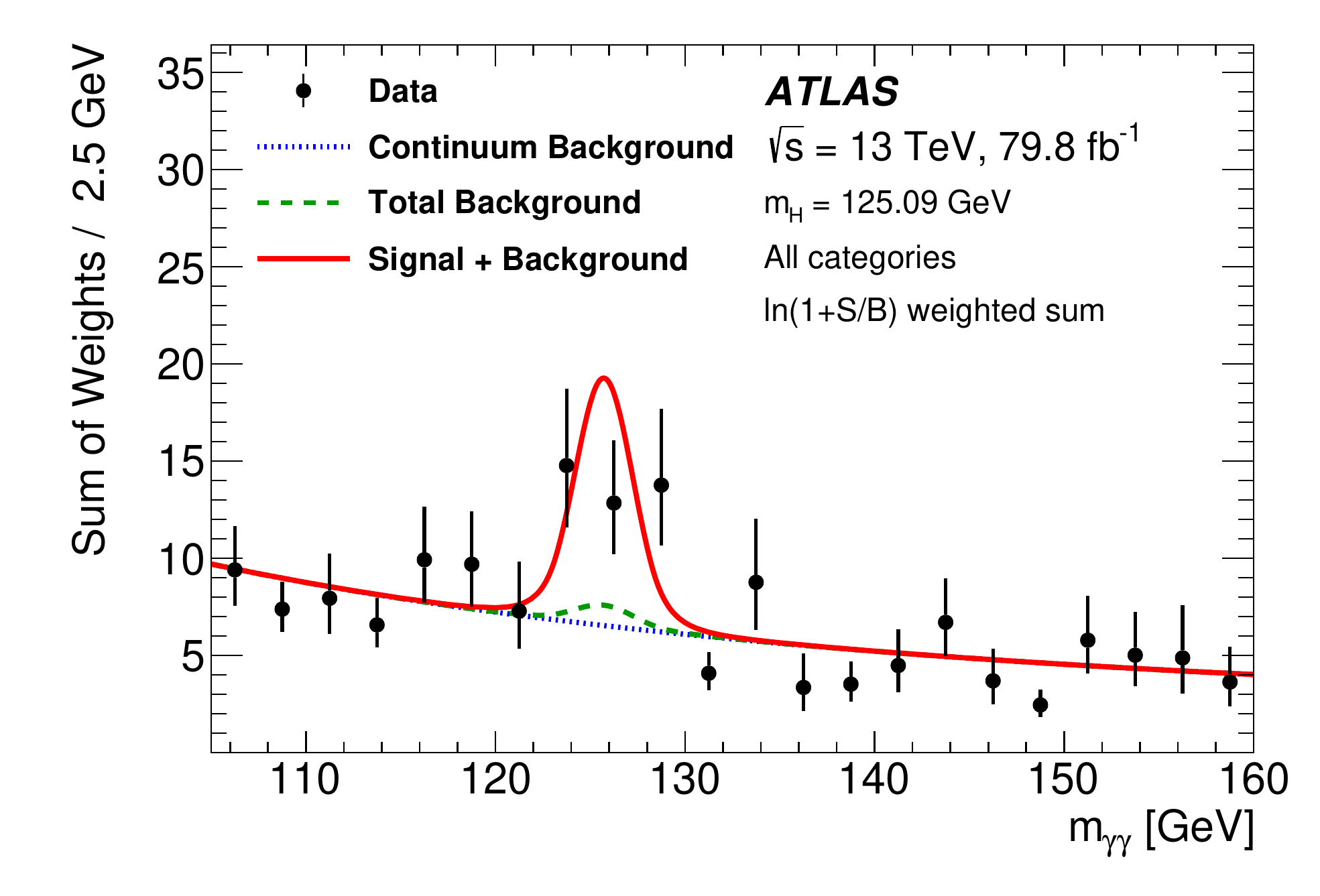}\label{fig:ttH_b}}
\end{minipage}
  \caption{Distributions of (a) the multivariate discriminant in the two same-sign $\mu$ channel for the
$t\bar{t}H$ CMS analysis~\cite{CMS-tth-ML} and (b) of the $\gamma\gamma$ invariant mass distribution in the $t\bar{t}H, H \rightarrow \gamma\gamma$ ATLAS analysis~\cite{ATLAS-tth-comb}.}
\label{fig:ttH}
\end{figure}

This production mode leads to a final state with $bbWW$ from the top quark decays to which the Higgs boson
decay products are added. Different analyses are performed to target the different Higgs boson decay modes.
\begin{itemize}
\item $H \rightarrow b\bar{b}$ : the final state contains 4~$b$-quarks. Hadronic, semi-leptonic or fully leptonic
decays of the top quarks are selected. The main background arises from the processes $t\bar{t}b\bar{b}$ and
$t\bar{t}c\bar{c}$ which are difficult to model. This channel also suffers from combinatorial background. Events
are classified according to the number of jets and the $b$-tagging properties of the jets. Multivariate discriminants
based on kinematical variables are trained for each category to enhance signal and background separation. The signal
yield is extracted from a global fit with the $t\bar{t}b\bar{b}$ and $t\bar{t}c\bar{c}$ background normalizations free
in the fit. An excess over the background-only hypothesis is observed with a significance of 1.4 (ATLAS~\cite{ATLAS-tth-bb}) 
and 1.6
(CMS~\cite{CMS-tth-bb,CMS-tth-bb2}) standard deviations with the 2015-2016 dataset, in agreement with the expected sensitivity for the SM
$t\bar{t}H$ production.
\item $H \rightarrow WW$ or  $H \rightarrow \tau\tau$ with multileptons final state. Final states with two leptons
(e or $\mu$) of same sign (one from the Higgs boson decay and one from one of the top quark decays), 
or three or four leptons (including up to one hadronic $\tau$ decay) are considered.
Opposite sign two lepton events are not considered to avoid the large background from $t\bar{t}$ production. 
The main irreducible backgrounds are the associated production of a vector boson $W$ or $Z$ with a  $t\bar{t}$ pair
and the diboson production processes. These backgrounds are estimated from simulation, normalized to NLO cross-section computations and verified
using data control regions. This channel also suffers from significant reducible backgrounds with non-prompt
leptons (mostly from $b$-hadron decays) or with mis-measured lepton charge. These reducible backgrounds are estimated directly from data.
Multivariate discriminants are trained
to separate the signal from the different background sources. An example is shown in Figure~\ref{fig:ttH}(a) for the
same sign 2$\mu$ final state in CMS. The main systematic uncertainties are related to the background modelling and
the total systematic uncertainty is comparable to the statistical uncertainty obtained with the 2015-2016 dataset.
The expected significance over the background-only hypothesis is 2.8 standard deviations. The observed excesses
correspond to significances of 4.1 (ATLAS~\cite{ATLAS-tth-ML}) 
and 3.8 (CMS~\cite{CMS-tth-ML}) standard deviations. The measured signal yields are
consistent with the SM expectations within approximately one standard deviation.
\item $H \rightarrow \gamma\gamma$. This decay mode provides a clean mass peak but suffers from low event
rate. The main irreducible background is $t\bar{t}\gamma\gamma$ production with other backgrounds coming from
$\gamma\gamma$ pairs not associated to $t\bar{t}$ , from non-prompt photon production and also
from other $H$ production modes. The ATLAS analysis~\cite{ATLAS-tth-comb}, performed with the 2015-2017 dataset,
separates events in all hadronic and leptonic regions. In each region, a neural network discriminant is
trained using a mixture of data and simulation to define categories with varying signal over background ratio.
In the most sensitive categories, the signal over background is around one for $\gamma\gamma$ masses close to the
Higgs boson mass. The signal is extracted from signal plus background fits to the  $\gamma\gamma$ mass distribution
with the background constrained by the mass sidebands. The observed (expected) significance is 4.1 (3.7)
standard deviations. Figure~\ref{fig:ttH}(b) shows the combined $\gamma\gamma$ mass distribution
where the mass peak is clearly visible above the background. A similar search is performed by the CMS collaboration
using the 2015-2016 dataset~\cite{CMS-hgg}.
\item  $H \rightarrow 4l$. This final state is clean but suffers from a low expected signal rate (less than one event).
It is used in the final combination although its expected sensitivity is significantly smaller than the channels
discussed above.
\end{itemize}
Combining the different decay modes above, and also with the run 1 data, the observed significance for the $t\bar{t}H$ observation
is 6.3 (5.2) standard deviations for the ATLAS~\cite{ATLAS-tth-comb}
(CMS~\cite{CMS-tth-comb}) analysis, in agreement with the expected significances (5.1 and 4.2
respectively). The ATLAS analysis sensitivity is higher because 2017 data are used for the $\gamma\gamma$ and $4\ell$
decay channels. The observed yield is $1.32\pm0.27$ ($1.26^{+0.31}_{-0.26}$) times the SM predictions from the ATLAS (CMS)
analysis.

\section{Observation of the decay $H \rightarrow b\bar{b}$}

Despite that $H \rightarrow b\bar{b}$ is the dominant Higgs boson decay mode, its observation
is difficult at the LHC, due to large background contamination.
The most sensitive channel is based on the $VH$ associated production, but $t\bar{t}H$, VBF~\cite{ATLAS-VBF-bb} and 
gluon fusion at high Higgs boson transverse momentum~\cite{CMS-gg-bb} productions are also investigated.

\begin{figure}[htb]
\centering
  \begin{minipage}{0.45\textwidth}
  \subfloat[]{\includegraphics[scale=0.34]{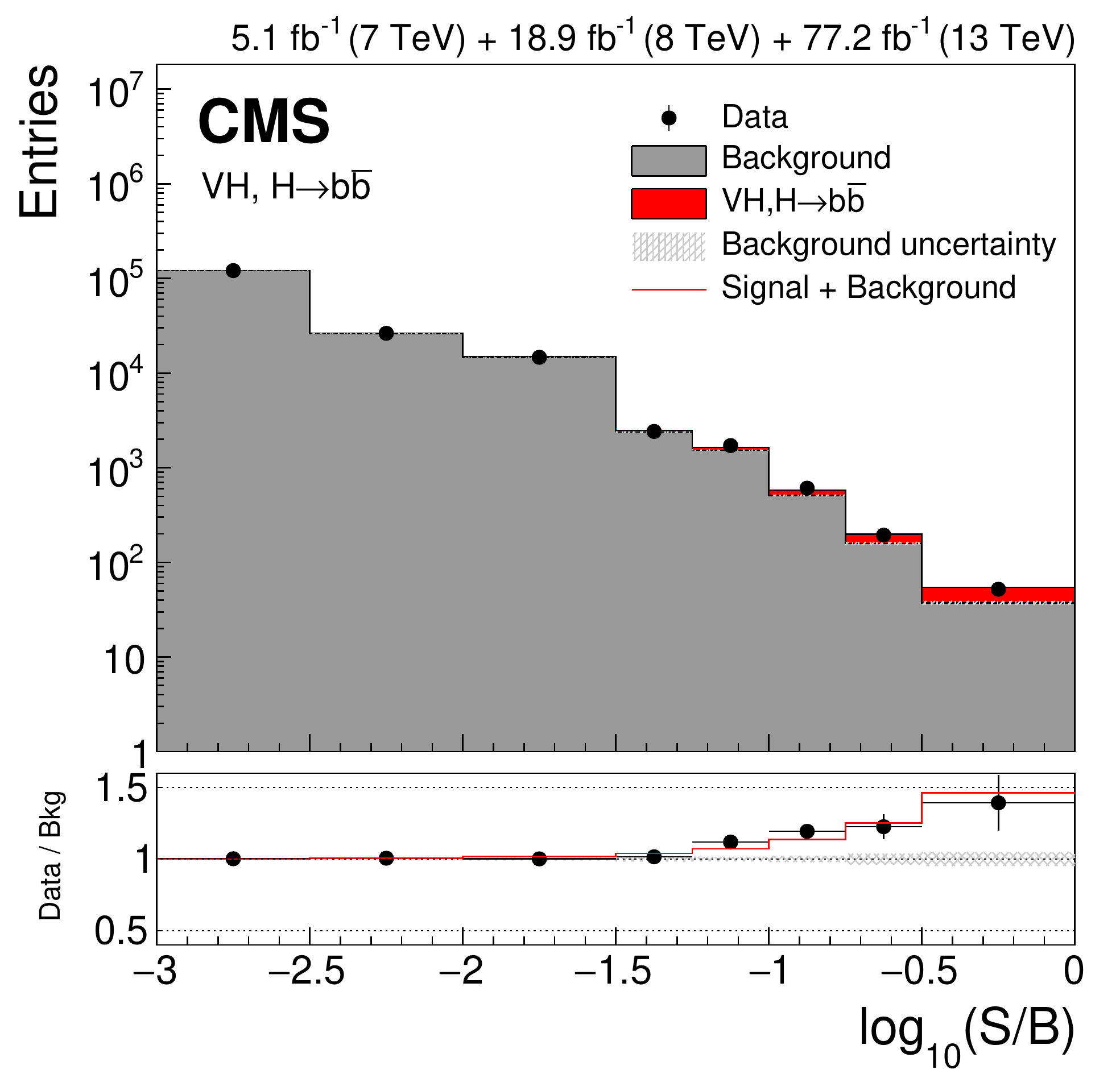}\label{fig:VHbb_a}}
  \end{minipage}
  \begin{minipage}{0.45\textwidth}
  \subfloat[]{\includegraphics[scale=0.38]{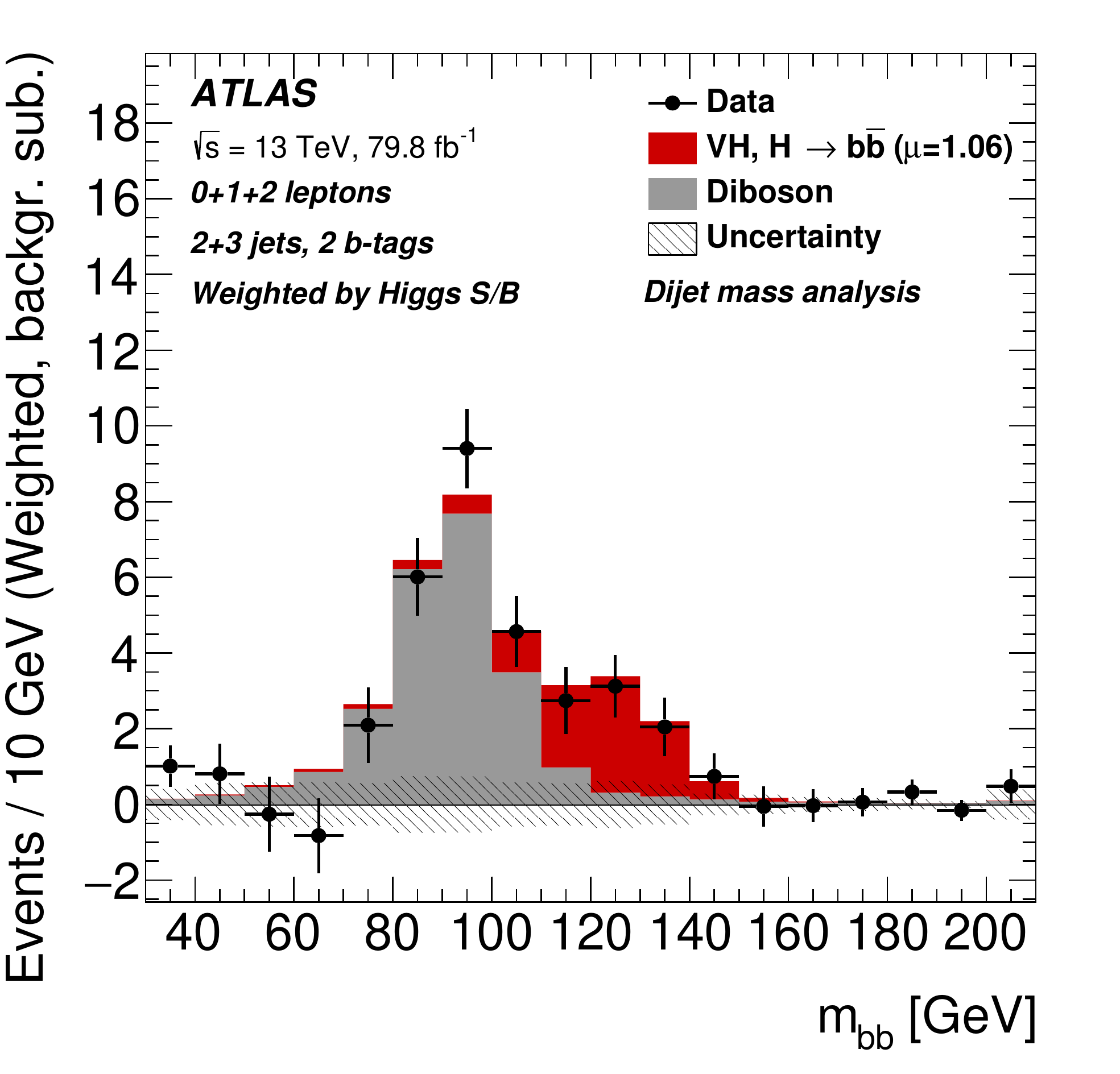}\label{fig:VHbb_b}}
   \end{minipage}
  \caption{Distributions of (a) the signal over background ratio of the events selected in the CMS $VH, H \rightarrow b\bar{b}$
analysis~\cite{CMS-VHbb} and (b) of the $b\bar{b}$ invariant mass for the cut-based ATLAS $VH, H \rightarrow b\bar{b}$
analysis~\cite{ATLAS-VHbb}.}
\label{fig:VHbb}
\end{figure}

The search of $H \rightarrow b\bar{b}$ with the $VH$
production mode separates events according to the number of leptons (electrons or muons). Events with
0 leptons (but with missing transverse momentum) and 2 leptons target the $ZH$ production while events with
one charged lepton target $WH$ production. Two $b$-tagged jets are required.
The main backgrounds arise from $V$+jets (mostly $Vb\bar{b}$) and $t\bar{t}$ productions. In addition to
$b$-tagging abilities, the main handles to separate signal and background are the $b\bar{b}$ invariant
mass and kinematical variables like the transverse momentum of the $V$ boson. Dedicated corrections are
applied to the $b$-jet energy to optimize the invariant mass resolution. A multivariate discriminant is trained
to separate signal from background in the different analysis categories. Control regions are also defined
to constrain the normalization of the $V$+jets and  $t\bar{t}$ backgrounds. The small background from multijet
events is derived directly from data. The main uncertainties are statistical uncertainties (data and MC sample size),
background modelling uncertainties, uncertainties in $b$-tagging performance and in the jet energy scale.
An important validation step is the search for the $VZ$ production followed by $Z \rightarrow b\bar{b}$ decay, which looks
like the signal except for the lower $b\bar{b}$ invariant mass. This process is observed at a rate consistent
with SM expectations with a 20\% accuracy. The expected sensitivity with the combined run 1 and 2015-2017 dataset to
the $VH, H \rightarrow b\bar{b}$ production is 5.1 (4.8) standard deviations for the ATLAS~\cite{ATLAS-VHbb} 
(CMS~\cite{CMS-VHbb}) analysis. The observed significance
is 4.9 (4.8) for ATLAS (CMS).
Figure~\ref{fig:VHbb}(a) shows the distribution of all events entering in the CMS analysis ordered by their
signal over background ratio. An excess of events at large values of signal over background is clearly
seen over the background-only hypothesis, consistent with the expected signal from $VH, H \rightarrow b\bar{b}$.
As cross-check, a cut-based analysis is also performed. The  $b\bar{b}$ invariant mass can then be investigated
as the final discriminating variable. The result of this analysis is shown in Figure~\ref{fig:VHbb}(b) for the
ATLAS analysis after all backgrounds but $VZ$ production are subtracted. The $VZ$ contribution, peaking
at a mass near 90~GeV, is clearly visible. The $VH$ production gives a significant excess of events at larger
values of the $b\bar{b}$ invariant mass.

After combining the $VH$ search with the searches in the other production modes, the $b\bar{b}$ decay
mode is observed at a significance of 5.4 (5.6) standard deviations by the ATLAS (CMS) Collaboration. The yield of events
compared to the SM predictions (assuming that the ratio of the different production mode is like in the SM)
is $1.01\pm0.20$ ($1.04\pm0.20$) for the ATLAS (CMS) analysis, in excellent agreement with the expectations.
The $VH, H \rightarrow b\bar{b}$ search, after combination with the other searches for $VH$ production in different
decay modes, also provides an observation of the $VH$ production.

\section{Decays to leptons}

\subsection{Observation of $H \rightarrow \tau\tau$}

$H \rightarrow \tau\tau$ decays can be searched for using either leptonic or hadronic decay modes of the $\tau$ lepton.
The invariant mass can be estimated using the missing transverse momentum measurement. The main background
arises from the $Z/\gamma^* \rightarrow \tau\tau$ process, estimated from simulation normalized
in a dedicated control region. The simulation is also extensively 
validated using for instance a large sample of $Z/\gamma^* \rightarrow \mu\mu$ decays. 
Backgrounds from fake hadronic $\tau$ candidates are derived from data.
Categories are defined
according to the Higgs boson production, the most sensitive ones are the category targeting VBF production
and the category targeting boosted $H$ production via gluon fusion.
The signal yield per production mode is extracted from fits of the $\tau\tau$ invariant mass distribution.
With the 2015-2016 dataset, the ATLAS~\cite{ATLAS-tautau} and CMS~\cite{CMS-tautau} experiments
observe this decay mode at more than five standard deviations
significance each. The total cross section (assuming the SM fractions for each production mode)
in the $H\rightarrow\tau\tau$ decay channel is measured with an accuracy
around 25\% in each experiment and found to be consistent with the SM prediction.
The ATLAS measurement for the cross-section times branching ratio to $\tau\tau$ pair is
$3.71 \pm 0.59 (\mathrm{stat})^{+0.87}_{-0.74} (\mathrm{syst})$~pb.
CMS reports a measurement of the the cross-section times branching ratio
of $1.09^{+0.27}_{-0.26}$ times the SM expectation.
Separate cross-section
measurements for the gluon fusion and VBF production processes are also reported.

\subsection{Search for $H \rightarrow \mu\mu$}

This channel allows one to probe the coupling to second generation fermions. The predicted branching
ratio is about 0.02\%. This channel has a low signal over background ratio with the background being largely
dominated by the Drell-Yan process. Categories are defined targeting gluon fusion and VBF productions and also
according to the expected di-muon mass resolution. The best category in CMS has a di-muon mass resolution
of about 1.2~GeV. The expected sensitivity to the SM predicted rate is about one standard deviation.
The observed limit at 95\% level is 2.9 times the SM expectation in the CMS analysis~\cite{CMS-mumu} based on 2015-2016 data combined
with run 1, and 2.1 times the SM expectation in the ATLAS analysis~\cite{ATLAS-mumu} using the 2015-2017 dataset.

\section{Decays to bosons}

\subsection{$H \rightarrow WW^*$}
This decay mode is searched for in the dilepton final state, the e-$\mu$ channel being the most sensitive one.
The Higgs boson mass cannot be reconstructed because of the two undetected neutrinos. The signal rate is large
but the signal over background ratio is smaller than one, even after dedicated kinematical selections. 
The main backgrounds from $WW$, $t\bar{t}$ and $W\gamma$ productions are normalized in dedicated control regions.
Events are divided into categories sensitive to the different production modes. The signal yields relative
to the SM expectations are measured separately for gluon fusion and VBF production.
The ATLAS results~\cite{ATLAS-WW} are $12.6^{+1.0}_{-1.0}\mathrm{(stat)}^{+1.9}_{-1.8}\mathrm{(syst)}$~pb 
and $0.50^{+0.24}_{-0.23}\mathrm{(stat)}\pm0.18\mathrm{(syst)}$~pb for the cross-section
times branching ratio for the gluon fusion and the VBF productions, in agreement with SM expectations. 
The CMS results~\cite{CMS-WW} on the ratio between the measured cross-sections
and the SM predictions are $1.24^{+0.20}_{-0.26}$ for gluon fusion and $0.24^{+0.74}_{-0.24}$ for VBF production.

\subsection{$H \rightarrow ZZ^*$}
The experimental signature with four leptons (e or $\mu$) in the final state is clean. 
The combined branching ratio is however small and great care is made to optimize the reconstruction and identification
of low transverse momentum leptons, especially electrons. The background arises mostly from the irreducible $ZZ^*$
production estimated from simulation with smaller reducible backgrounds from $t\bar{t}$ and $Z$+jets processes with
non-prompt leptons.
Figure~\ref{fig:bosons}(a) shows the four-lepton mass distribution observed by ATLAS~\cite{ATLAS-ZZ} with the
2015-2017 dataset. The data are divided in several categories (seven in this analysis) to optimize the sensitivity
to the different production modes. 
The ATLAS analysis obtains a measurement of the ratio between the observed
and predicted yields of $1.19^{+0.16}_{-0.15}$ where the main contribution to the uncertainty is the data
statistical uncertainty.
The CMS analysis~\cite{CMS-ZZ} reports a signal yield relative to the SM prediction of $1.10^{+0.19}_{-0.17}$ with 
the 2017 dataset.

\subsection{$H \rightarrow \gamma\gamma$}
This channel offers a clear signature with a narrow mass peak over a smooth background. There is a large
background mostly from continuum diphoton production, with the inclusive signal over background ratio being
around few percents. For this search, the invariant mass resolution is optimized and studied starting
from samples of $Z \rightarrow ee$ events. Events are categorized according to the signal over background ratio
and to the invariant mass resolution as well as according to the production modes to enhance the sensitivity.
The overall signal yield compared to the SM prediction is measured to be
$1.06^{+0.14}_{-0.12}$  by the ATLAS analysis~\cite{ATLAS-phot} with the 2015-2017 dataset, while the CMS analysis~\cite{CMS-hgg}
obtains a value of $1.18^{+0.17}_{-0.14}$  with the 2015-2016 dataset.

\subsection{Higgs boson mass measurement}
The Higgs boson mass can be measured using the high resolution $ZZ^*$ and $\gamma\gamma$ final state.
Combining the measurements in these two channels from 2015-2016 data and from run 1, the ATLAS collaboration reports a value
of the Higgs boson mass of $124.97 \pm 0.24$~GeV~\cite{ATLAS-mass} (with $\pm0.19$ GeV of statistical uncertainty and $\pm0.13$~GeV
of systematic uncertainty, mainly from uncertainties in the photon energy scale).
With the  $ZZ^*$ channel from 2015-2016 data, the CMS collaboration reports a mass value of $125.26\pm0.21$~GeV~\cite{CMS-mass}. 
At the same time, a direct upper
limit on the decay width is set at 95\% confidence level at 1.1~GeV.
This is still far above the predicted width in the SM which is about 4~MeV. 
A more model dependent constrain on the Higgs boson width can be derived comparing the rate of $gg\rightarrow H^{(*)}\rightarrow ZZ^{(*)}$
events in the on-shell and off-shell Higgs mass regions. The ATLAS analysis with 2015-2016 data sets a model-dependent limit
at 14.4~MeV on the decay width, at 95\% confidence level~\cite{ATLAS-offshell}.

\subsection{Total and differential cross-section measurements with $ZZ^*$ and $\gamma\gamma$ final states}

The total cross-section can be measured with little model dependence using the $ZZ^*$ and $\gamma\gamma$ final states.
With the 2015-2016 dataset,
the ATLAS Collaboration reports a measurement of $57.0^{+6.0}_{-5.9}\mathrm{(stat)}^{+4.0}_{-3.3}\mathrm{(syst)}$~pb~\cite{ATLAS-xsec},
while the CMS analysis finds $61.1\pm6.0\mathrm{(stat)}\pm3.7\mathrm{(syst)}$~pb~\cite{CMS-xsec}. These measurements are in good agreement
with the SM expectation of $55.6\pm2.5$~pb. 

\begin{figure}[htb]
\centering
  \begin{minipage}{0.45\textwidth}
  \subfloat[]{\includegraphics[scale=0.32]{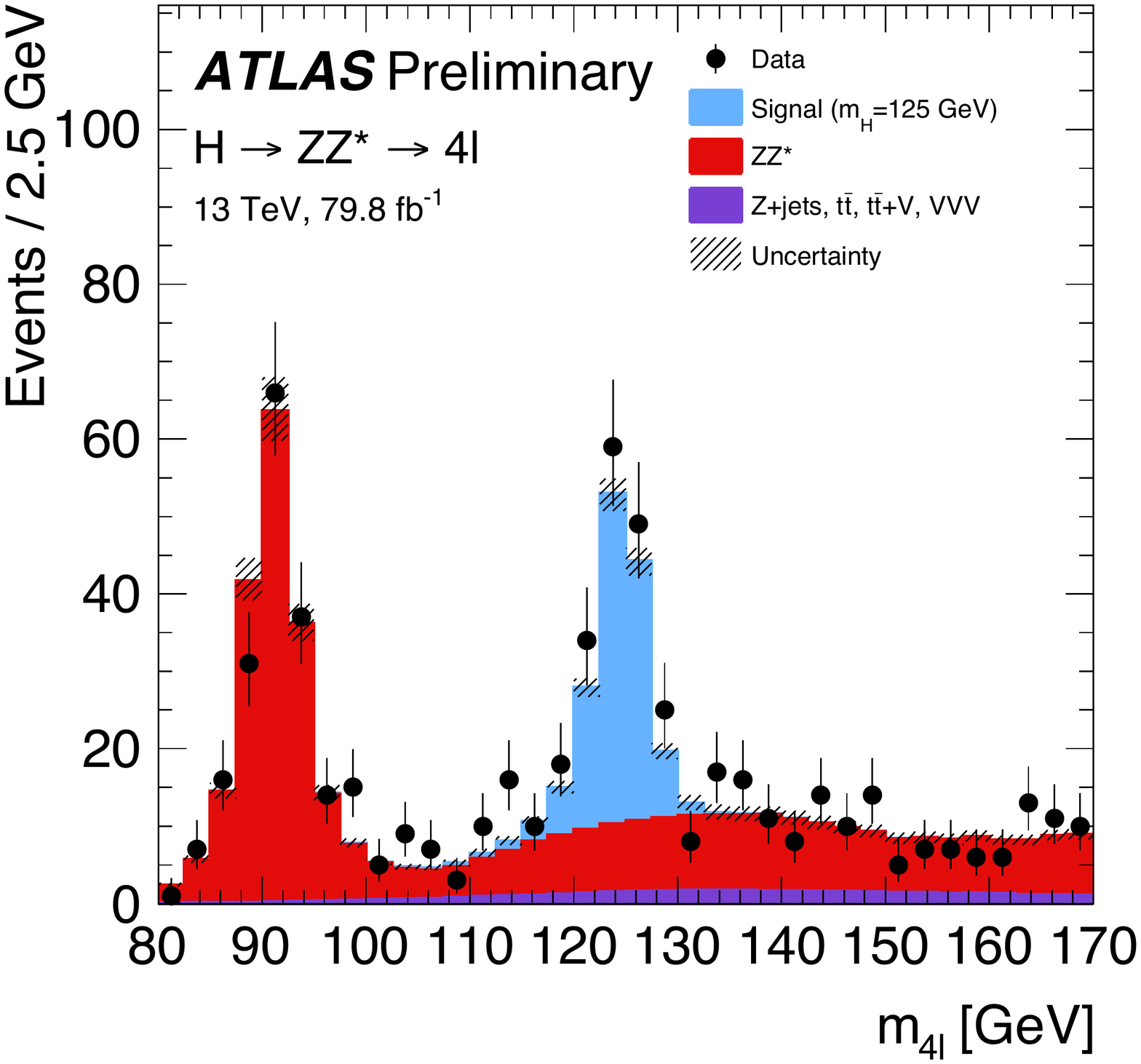}\label{fig:bosons_a}}
  \end{minipage}
  \begin{minipage}{0.45\textwidth}
  \subfloat[]{\includegraphics[scale=0.34]{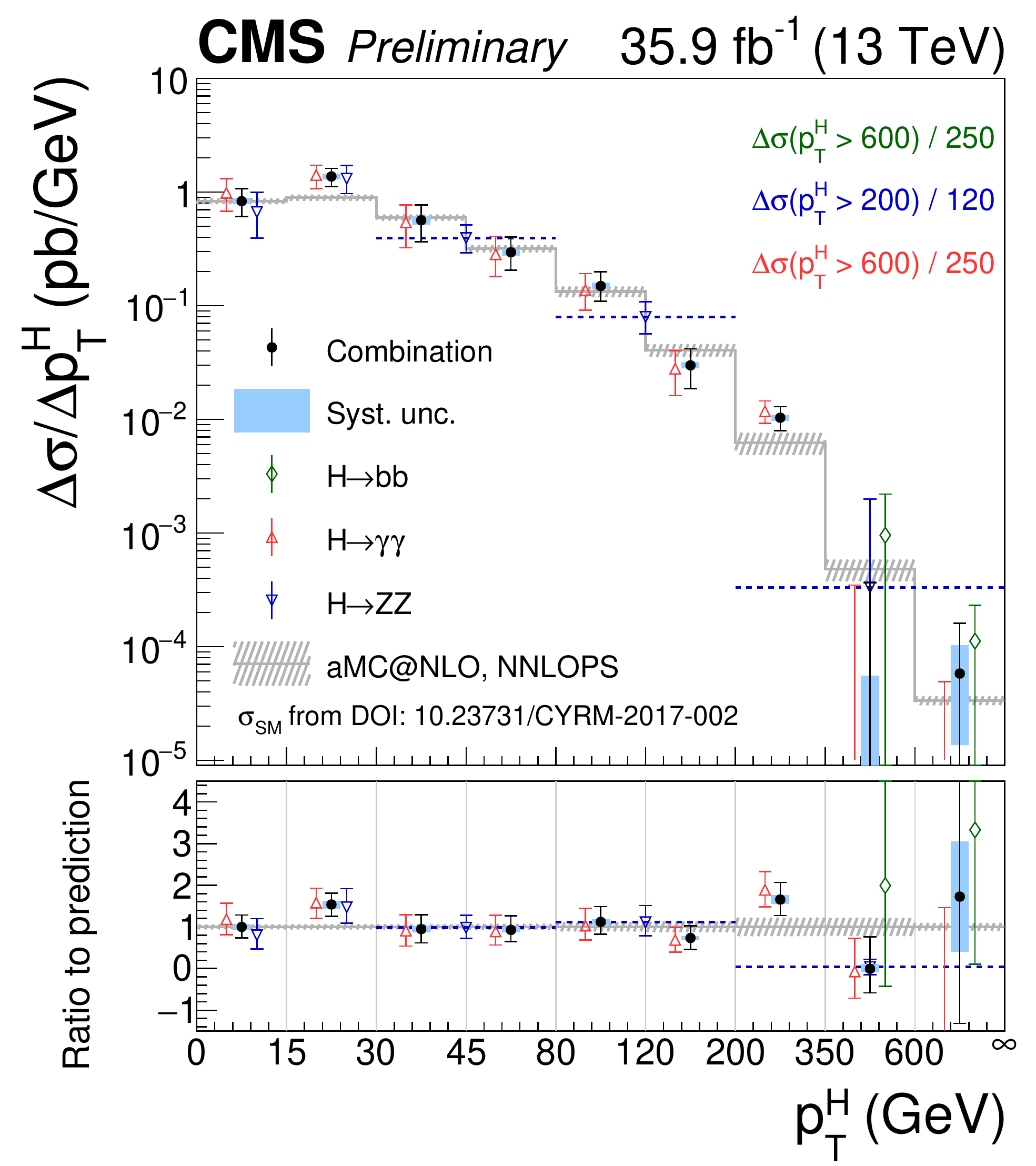}\label{fig:bosons_b}}
  \end{minipage}
\caption{(a) Distributions of the 4 lepton invariant mass in the $H\rightarrow ZZ^*$ channel from the
ATLAS experiment~\cite{ATLAS-ZZ} and (b) measurement of the differential cross-section as a function of the
Higgs boson transverse momentum with the CMS experiment~\cite{CMS-xsec}}.
 \label{fig:bosons}
\end{figure}

Differential cross-sections as a function of several variables can also be measured in these channels.
Figure~\ref{fig:bosons}(b) illustrates the measurement performed by CMS of the cross-section as a function of the
transverse momentum of the Higgs boson~\cite{CMS-xsec}. 
The accuracies of the $\gamma\gamma$ and $ZZ$ channels are comparable.
At high transverse momentum, the $b\bar{b}$ analysis~\cite{CMS-gg-bb} is also
used. The measured differential cross-section agrees with the state-of-the-art theoretical predictions over the
full transverse momentum range. Similar results are also reported by the ATLAS Collaboration~\cite{ATLAS-ZZ,ATLAS-phot}.

\section{Rare decay searches}
Rare decays predicted in the SM are also searched for.
The first example is the decay $H\rightarrow \ell\ell\gamma$ which offers a rich structure depending
on the dilepton invariant mass. With the 2015-2016 data, the CMS experiment reports a limit at the level of 8 times
the SM expectations for dilepton invariant mass close to the $Z$ boson mass~\cite{CMS-llgamma}. For dilepton masses significantly lower 
than the $Z$ boson mass (in which
case the process is dominated by a virtual photon), the limit is 4 times the SM expectations.
Searches for $H\rightarrow J/\psi \gamma$ decays (with $J/\psi \rightarrow \mu\mu$) can probe the coupling
of the Higgs boson to charm quarks. With the 2015-2016 data, the ATLAS experiment obtains a
branching ratio upper limit of $3.5\cdot10^{-4}$ at 95\% confidence level~\cite{ATLAS-jpsigamma}. This is a factor 120 times
larger than the SM prediction.
A search for $H\rightarrow c\bar{c}$ decay is also performed in the $VH$ associated production mode, with an 
upper limit on the yield around 100 times the predicted SM value~\cite{ATLAS-vhcc}.

\section{Constraints on the Higgs boson couplings}

Studies of the Higgs boson couplings to probe physics beyond the Standard Model
can be achieved combining information from all the investigated production and decay modes.
One framework used to report the Higgs coupling properties is the $\kappa$ framework in which the same
coupling structure as in the SM is assumed and the couplings are just scaled by coupling modifiers $\kappa_i$.
For channels involving loops, like the couplings between $H$ and photons or gluons, the coupling strength
can be either parameterized by an effective coupling scale factor or computed resolving the loop content with its
SM particle content. In the later case, the Higgs boson coupling to gluons is for instance mostly
driven by the $\kappa_{t}$ factor with a small contribution from $\kappa_{b}$ and their interference.
A detailed discussion of this framework and how the rates in each production mode and decay channel are
modified can be found in Ref.~\cite{ref-kappa}.
Since the total width of the Higgs boson is not directly accessible at the LHC, some assumptions are needed
to avoid the degeneracy between the $\kappa$ scaling and the total width.

The combinations reported here are based on 2015-2016 data for most decay modes with the 2015-2017 dataset used
for the ATLAS $H \rightarrow \gamma\gamma$ and $H \rightarrow ZZ^*$ channels, including the $t\bar{t}H$ production
mode categories of these analyses. The CMS results are described in Ref.~\cite{CMS-comb} and the ATLAS
results in Ref.~\cite{ATLAS-comb}.

\begin{figure}[htb]
\centering
  \begin{minipage}{0.45\textwidth}
  \subfloat[]{\includegraphics[scale=0.35]{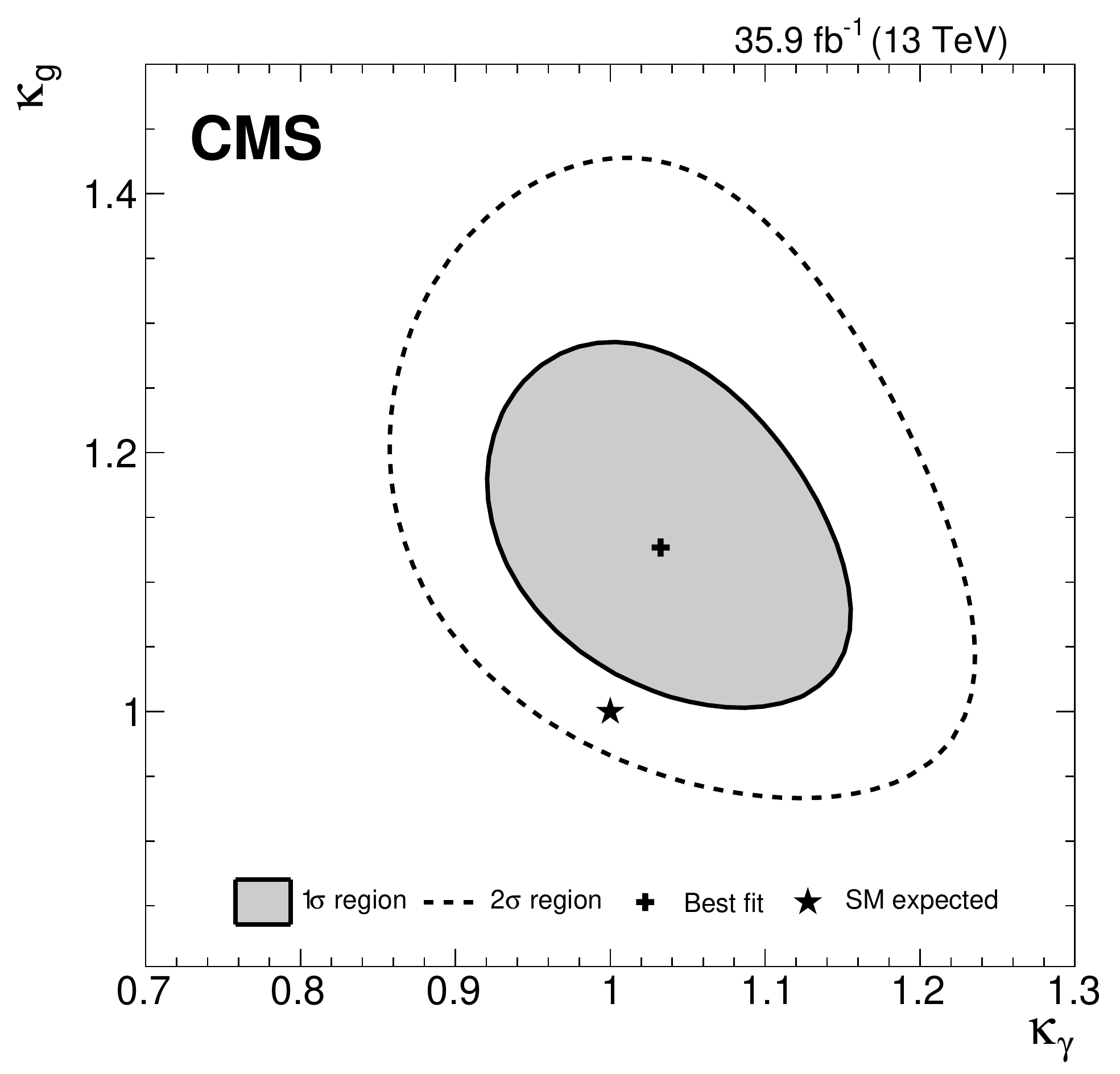}\label{fig:kappa_a}}
  \end{minipage}
  \begin{minipage}{0.45\textwidth}
  \subfloat[]{\includegraphics[scale=0.43]{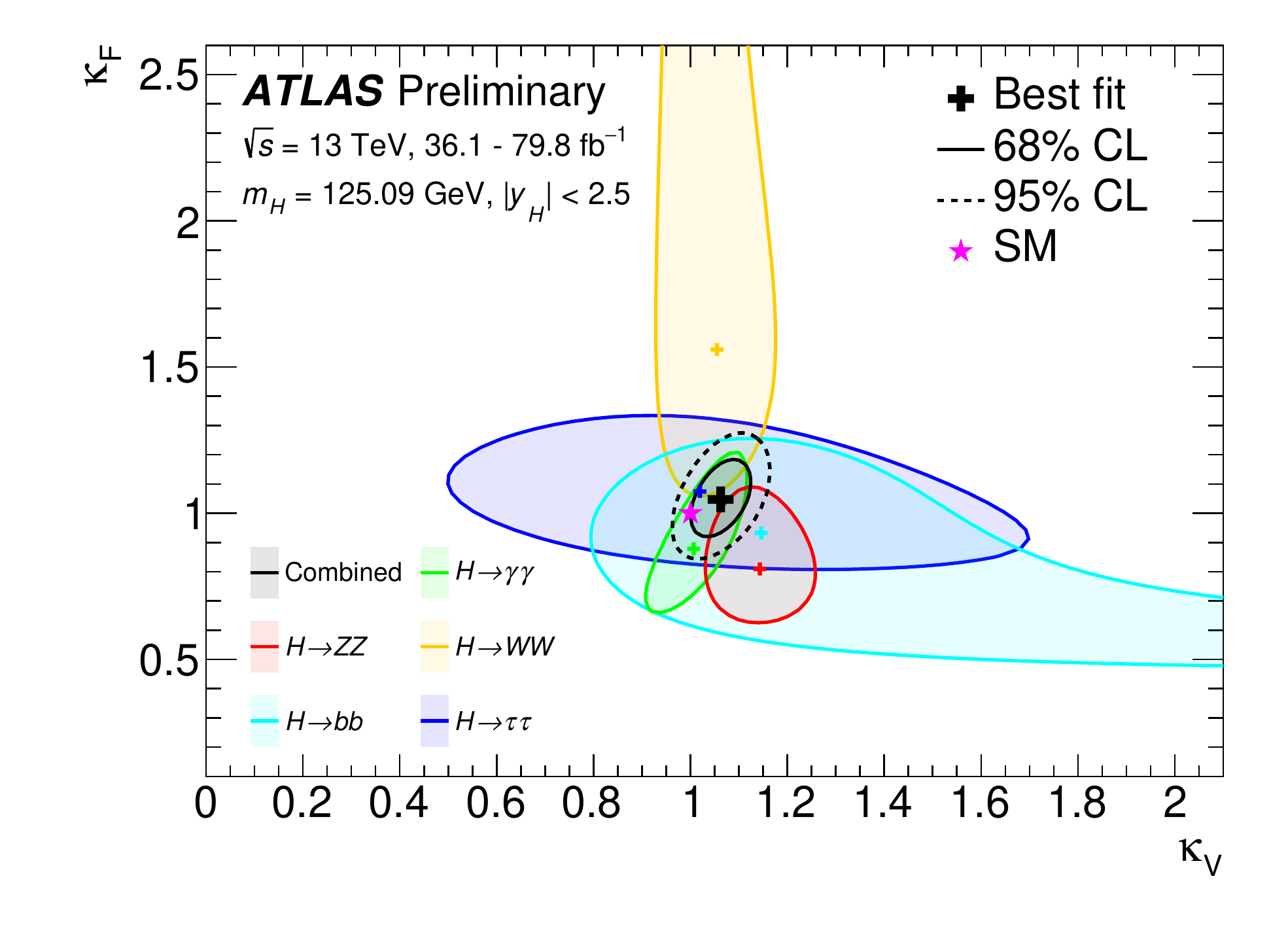}\label{fig:kappa_b}}
  \end{minipage}
 \caption{Measurements of (a) the effective couplings to photons and gluons in the CMS combined analysis~\cite{CMS-comb} and
 (b) of the effective couplings to fermions and bosons in the ATLAS combined analysis~\cite{ATLAS-comb}.}
  \label{fig:kappa}
\end{figure}

Figure~\ref{fig:kappa}(a) shows the constraints on the gluon and photon effective couplings from the CMS analyses assuming
all the other couplings are like predicted by the SM, thus probing new physics contributions to the loop mediated
couplings of the Higgs boson to gluon and photons. The values are found consistent with the SM expectation of one.
Figure~\ref{fig:kappa}(b) shows the constraints on the effective couplings to fermions and bosons from the ATLAS analyses. 
All $H$ bosons couplings
to fermions are assumed to scale by the same factor $\kappa_F$ and similarly for the couplings to bosons with the scale
factor $\kappa_V$. Loops are resolved assuming the SM content and no other new physics contribution. The total Higgs boson
width is also computed assuming rescaled SM contributions and no invisible or undetected decay modes. The data are consistent
with the SM predictions. This Figure also shows the constraints from each individual decay mode and how the combination
allows one to obtain more stringent constraints.

\begin{figure}[htb]
\centering
  \begin{minipage}{0.45\textwidth}
  \subfloat[]{\includegraphics[scale=0.41]{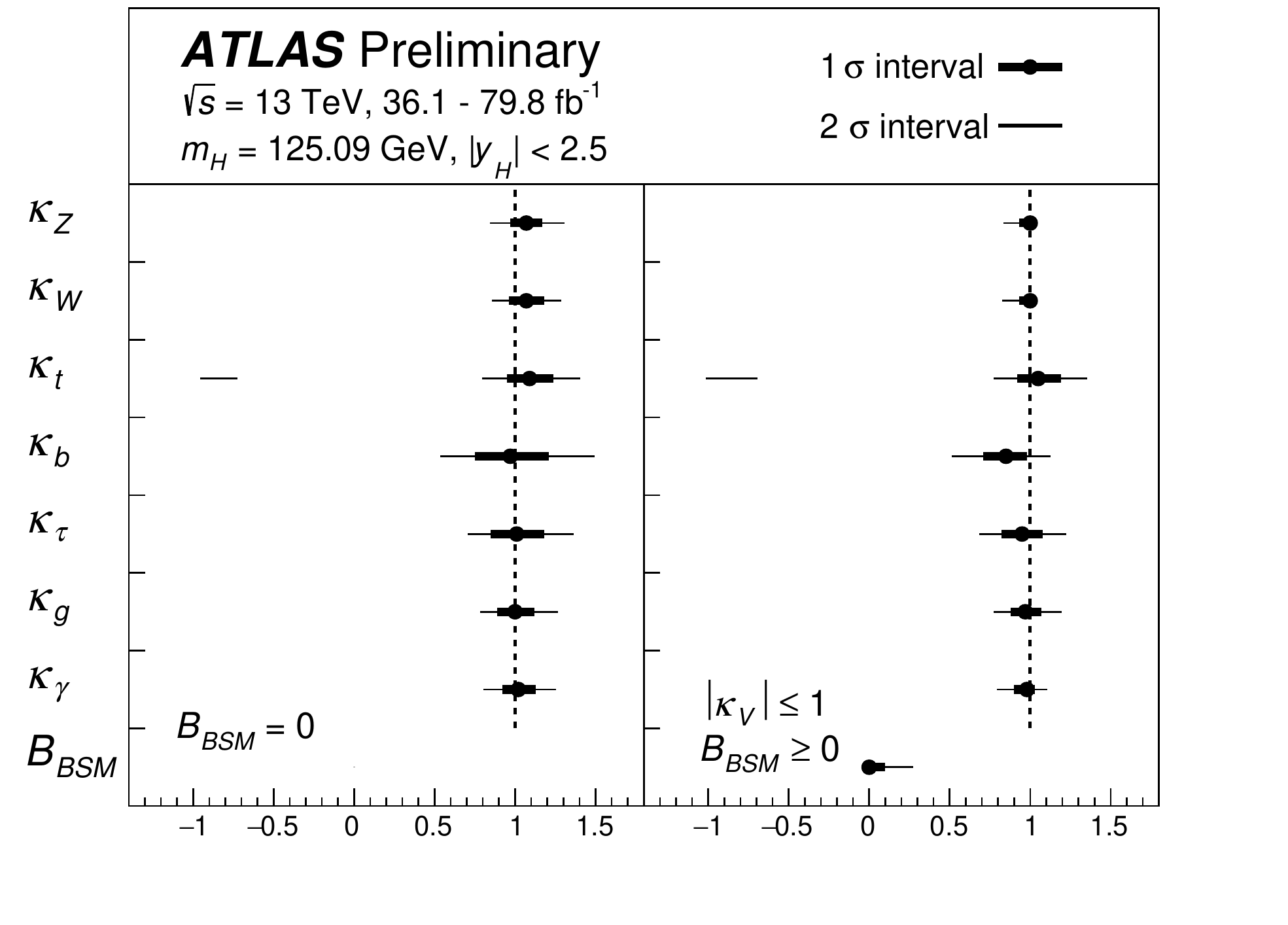}\label{fig:couplings_a}}
  \end{minipage}
  \begin{minipage}{0.45\textwidth}
  \subfloat[]{\includegraphics[scale=0.33]{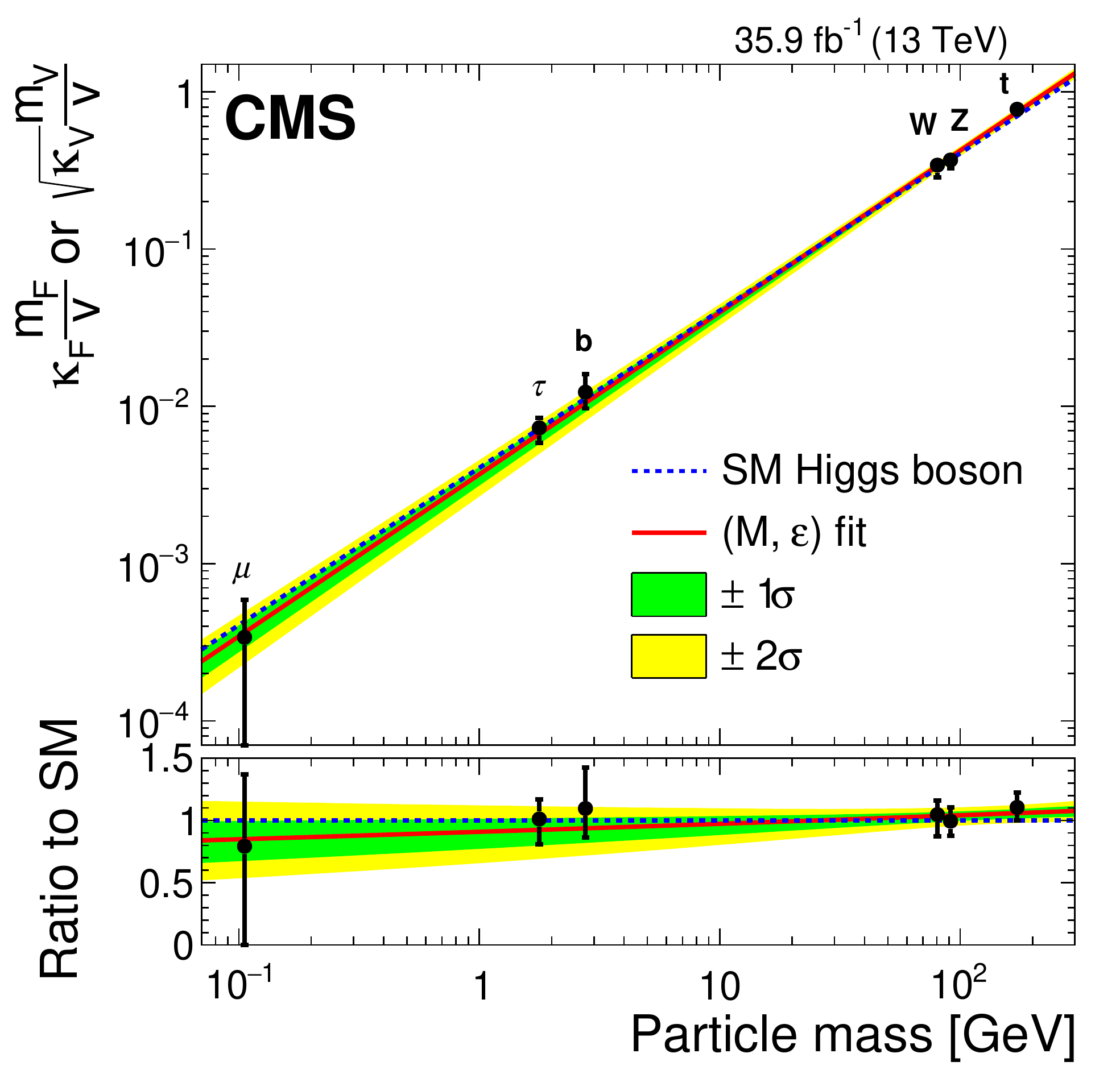}\label{fig:couplings_b}}
  \end{minipage}
 \caption{Measurements of (a) all $\kappa$ coupling modifier parameters simultaneously for two different assumptions
in the ATLAS combined analysis~\cite{ATLAS-comb} and (b) of the scaling of the Higgs boson couplings as a function of the particle
mass in the CMS analysis~\cite{CMS-comb}.}
\label{fig:couplings}
\end{figure}

Figure~\ref{fig:couplings}(a) shows the measurements of all $\kappa$ parameters achieved in the ATLAS analysis.
Two different assumptions are shown. In the first one, it is that there is no invisible or undetected $H$ decay mode. In the
second one, a beyond Standard Model contribution to the $H$ decay width is allowed but the parameter $\kappa_Z$ and
$\kappa_W$ are restricted to be less than one to break the degeneracy between coupling strengths and total width. This
assumption is natural in several extensions of the SM, like for instance two Higgs doublets models where the couplings
to $W$ and $Z$ bosons are shared between the two neutral scalar bosons. 
The measurement of the $b\bar{b}$ decay mode has an important impact on all results since this decay mode drives the total Higgs boson
width in the SM. In this case where effective couplings are used for photon and gluon couplings, the measurement
of the $t\bar{t}H$ production mode is the only input that allows to constrain $\kappa_t$. The measured values are in good
agreement with the SM predictions. The accuracy on the $\kappa$ parameters varies between 10\% and 20\% and is already
better than the combined run 1 results from ATLAS and CMS.
Similarly Figure~\ref{fig:couplings}(b) shows with similar assumptions how the rescaled couplings vary with the mass
of the particle. The observed scaling agrees well with the SM predictions. 

The ATLAS and CMS collaborations also report results~\cite{ATLAS-ZZ,ATLAS-phot,ATLAS-STX,CMS-comb} using the simplified template cross-section framework~\cite{cross-sections}.

\FloatBarrier

\section{Conclusions}
Higgs boson measurements based on 35 to 80~fb$^{-1}$ of proton-proton collision data recorded at the LHC by the
ATLAS and CMS experiments have been reviewed. With this dataset, important milestones for Higgs boson physics at the
LHC have been reached with the observation of the $t\bar{t}H$ production and of the decay $H \rightarrow b\bar{b}$.
The four main production processes and five main decay modes of the Higgs boson are now established. In addition,
measurements involving bosons in the final state reach higher precision allowing quasi-model independent measurements
of differential cross-sections. Studies of the Higgs boson couplings using the combination of all investigated production
and decay modes are reported with accuracy on coupling modifier parameters reaching 10 to 20\%. The results
are consistent with the Standard Model expectations.

\end{document}